# Data-Driven Web-Based Patching Management Tool Using Multi-Sensor Pavement Structure Measurements

**Sneha Jha**
Department of Agriculture and Biological Engineering
Purdue University, West Lafayette, IN 47907, USA.
Email: jha16@purdue.edu

**Yaguang Zhang**
School of Electrical and  Computer Engineering,
Purdue University, West  Lafayette, IN 47907, USA
Email: zhan1472@purdue.edu

**Bongsuk Park**
Lyles School of Civil Engineering,
Purdue University, West Lafayette, IN 47907, USA.
E-mail: bongsuk@purdue.edu
ORCID: 0000-0003-4606-4408

**Seonghwan Cho**
Pavement Research Engineer
Indiana Department of Transportation, West Lafayette, IN 47906, USA
E-mail: scho@indot.in.gov

**James V. Krogmeier**
School of Electrical and  Computer Engineering,
Purdue University, West  Lafayette, IN 47907, USA
E-mail: jvk@purdue.edu

**Tandra Bagchi**
Lyles School of Civil Engineering,
Purdue University, West Lafayette, IN 47907, USA.
E-mail: tbagchi@purdue.edu

**John E. Haddock**
Lyles School of Civil Engineering,
Purdue University, West Lafayette, IN 47907, USA.
E-mail: jhaddock@purdue.edu

## ABSTRACT

Automating pavement maintenance suggestions is challenging, especially for actionable recommendations such as patching location, depth and priority. It is common practice among State agencies to manually inspect road segments of interest and decide maintenance requirements based on the pavement condition index (PCI). However, standalone PCI only evaluates the pavement surface condition and coupled with the variability in human perception of pavement distress, limits the accuracy and quality of current pavement maintenance practices. Here, a need for multi-sensor data integrated with standardized pavement distress condition ratings is required. This study explores the possibility of estimating the appropriate pavement patching strategy (i.e., patching location, depth, and quantity) by integrating pavement structural and surface condition assessment with pavement specific ratings of distress. Especially, it combines pavement structural condition assessment parameter; falling weight deflectometer deflections along with surface condition assessment parameters; international roughness index, and cracking density for a better representation of overall pavement distress condition. Then, a pavement specific threshold-based patching suggestion algorithm is implemented to evaluate the pavement overall distress condition into a priority-based patching suggestion. The novelty in the use of pavement specific thresholds is placed on its data-driven ability to determine threshold values from current road condition measurements using a reliability concept validated by the theoretical pavement condition rating, pavement structural number. A web-based patching manager tool (PMT) was implemented to automate the patching suggestion procedure and visualize the results. Validated with road surface images obtained from three-dimensional laser sensors, PMT could successfully capture localized distresses in existing pavements.

## INTRODUCTION

Road networks are one of the largest continuous man-made structures in active usage and constitute a major portion of economic activities in the modern world. With increased traffic loads, road maintenance has only ever increased in scale. The Indiana Department of Transportation (INDOT), which maintains about 18000 centerline kilometers of roads, invest billions of public funded dollars on maintenance initiatives. Most state agencies rely on the experiences of scoping engineers and experts to determine pavement maintenance strategies based on various pavement condition indices. However, the scale of maintenance and variability in human perception of pavement distress limits the implementation and maintenance strategies. To cover large areas of interest with better accuracy, automated pavement maintenance suggestion systems are in demand *(1, 2)*. Pavement maintenance requirements involve more than the detection of localized distresses such as potholes, cracking, faulting, because they must also account for various environmental factors and/or traffic loading *(3)*. Instead, pavement maintenance is grouped into three categories, preservation, rehabilitation, and reconstruction based on low, medium, or high level of pavement distress severity. The classification of pavement condition into these categories are based on thresholds. These thresholds are used to categorize the pavement into preservative, rehabilitative or reconstructive maintenance requirement categories, but are often decided based on experience, human resources, time, and budget *(3-7)*.

Surface patching, thin overlay, and structural overlay are typical methods used for preservation and rehabilitation of pavements. These techniques use either visual distress indicators or decision trees based on the pavement condition indices such as ride quality, pavement distress index etc. *(8-10)*. PCI includes international roughness index (IRI), rut depth, and crack density (CD) to classify pavement surface conditions *(11-13)*. Vehicle-mounted IRI is an especially popular method of pavement assessment due to its ease of measurement and high resolution over long distances. Even though PCI is successful in characterizing pavement rideability, it does not properly account the effect of pavement structural conditions.

Typically, the relation between surface and structural conditions of the road is probabilistic, yet researchers have not been able to correlate the PCI and structural condition measurements *(14, 15)*. The structural condition of a pavement is most often measured by the falling weight deflectometer (FWD) and structural number index calculated from the FWD center deflection *(16, 17)*. Although Sollazzo et al. *(18)* have used neural networks and machine learning techniques to estimate structural number index from IRI to predict pavement condition, they do not yet relate to overall structural condition. The overall structural condition of pavement is better represented by including the deflection basin parameters (DBPs), which reflect the layer condition at different depths in the pavement *(19, 20)*. Comprehensive pavement maintenance requires both surface and structural condition assessments.

Thus, an information fusion of the surface and structural condition parameters is desirable to generate comprehensive pavement condition assessment. Integrating the IRI, CD and FWD measurement data spatially, can allow both, local distress detection using high resolution surface condition parameters and overall road structural condition using the FWD DBPs. On these grounds, a complete patching suggestion based on multi-sensor data fusion approach is proposed in this study.

## Objective and Scope

The primary objective of this study is to develop an enhanced patching suggestion tool by incorporating both functional and structural conditions of existing pavements into a threshold-based patching suggestion algorithm. Furthermore, this study proposes a methodology to determine the threshold values of FWD, IRI, and CD parameters using real-life pavement condition measurements and local road conditions. As a first step, this study focuses on full-depth asphalt pavements, a typical pavement type in Indiana. Field FWD, IRI, and CD data were collected from the INDOT and include all road classifications of full-depth asphalt pavements, i.e., State Road, US-Highway, and Interstate Highway. The collected field data were used to determine appropriate threshold values to build the patching suggestion algorithm and to implement the patching suggestions in the web-based application; Patching Management Tool (PMT) to aid in visualization and scoping of road conditions over large area networks.

## DESCRIPTION OF PAVEMENT CONDITION RATING (PCR) PARAMETERS

The multi-sensor data fusion method for creating high resolution patching suggestion spatially integrates pavement condition rating (PCR) parameters, IRI and CD, measured from the three-dimensional (3D) laser imaging sensor with pavement deflection basin parameters collected from FWD testing.

The IRI data used in this study was processed by a commercial software, WayLink ADA3 *(21, 22)*. This software can recalculate IRI for selected spatial resolutions and match the results with the road surface images obtained from 3D laser imaging sensor and right of way (ROW) images. For the purpose of this study, 1.8 m length along the road was selected as the spatial resolution. The WayLink ADA3 software also implements an improved crack detection technique from 3D laser images using image processing and deep-learning neural networks *(23)*. The crack detection results provide a percentage of cracking on left wheel path and right wheel path with the same spatial resolution as IRI.

FWD deflection parameters used in this study include two deflection values (D0 and D60) and three DBPs, which are used to evaluate the structural conditions of asphalt pavements *(24)*. D0, the maximum deflection measured at the center of FWD loading plate, is usually used to assess the overall pavement structural conditions, while D60, deflection measured 1500 mm away from the loading plate, represents the subgrade condition. The three DBPs are surface curvature index (SCI), base damage index (BDI), and base curvature index (BCI). SCI is defined as the difference between D0 and D12 (deflection measured 300 mm away from the loading plate) and used to characterize the structural conditions of pavement upper layers. In addition, BDI and BCI are used to indicate the structural conditions of base and subbase layers, respectively. BDI is the difference between D12 and D24 (deflection measured 600 mm away from the loading plate), and BCI is calculated by subtracting D36 (deflection measured 900 mm away from the loading plate) from D24. The use of DBPs as structural indicators allows to determine more specific and appropriate patching depth by interpreting the structural conditions of each layer in full-depth asphalt pavements. **Figures 1a** and **1b** show examples of PCR parameters (IRI and D0) as visualized in the PMT web application interface. In addition, **Figures 1c and 1d** show the road surface images and ROW images for each distance measurement instrument (DMI) number with 1.8m spatial resolution obtained from the WayLink ADA3 software.

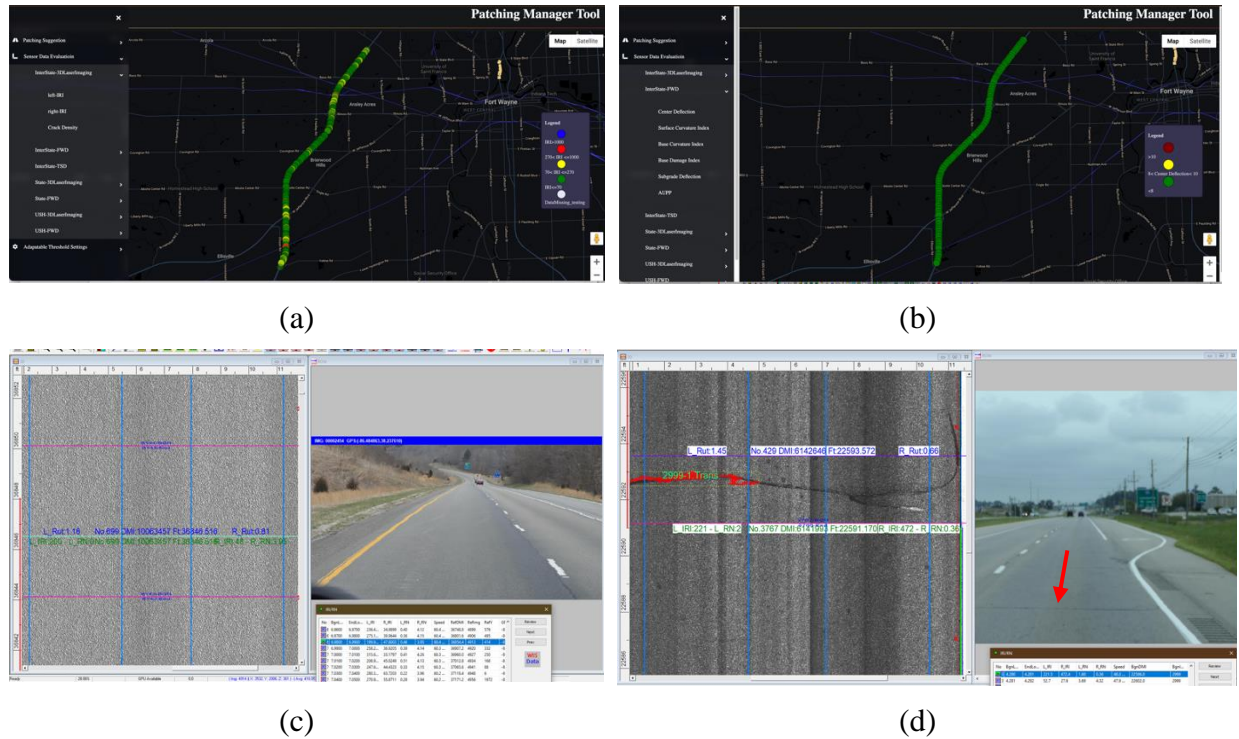

(a)          (b)

(c)          (d)

**Figure 1 Example of visualized PCR: (a) IRI, (b) D0, (c) road surface images and ROW images at the IRI measurement, and (d) road surface images and ROW images of transverse cracking**

## DETERMINATION OF THRESHOLDS FOR PAVEMENT CONDITION RATING PARAMETERS

### Determination of FWD Parameters Thresholds

Thresholds of FWD parameters were preliminarily determined based on the concept of design reliability, which is the probability of good pavement sections over the design period. According to the Mechanistic-Empirical Pavement Design Guide, the reliability concept is recommended for determining performance threshold values to achieve a balance between maintenance costs and overall pavement structural health *(25)*. In this study, a reliability analysis was conducted using the INDOT FWD database to determine preliminary thresholds of FWD parameters. Since threshold values determined by the reliability concept are dependent on the current conditions of local pavement systems, a further verification is needed to determine whether the reliability-based approach can provide reasonable thresholds in the aspect of structural capacity of pavements. Therefore, the structural number ratio (SNR), the most common pavement structural condition index, was used to further verify the preliminary threshold values (reliability-based thresholds). This section describes the details of FWD parameters thresholds determination, following two steps: (i) determine the preliminary thresholds using the reliability concept, and (ii) verify the preliminary thresholds using the SNR concept.

*Determination of Preliminary FWD Thresholds Using Reliability*
Two threshold values were determined for each FWD parameter to define three levels of PCR as good, fair, and poor structural conditions. For each road classification, two reliability levels were selected with 5% interval, based on the recommended level of reliability in the AASHTO 1993 design guide *(26)*. Interstate Highway used 90% and 95% for lower and upper limits, respectively. US-Highway used 85% to 90% of reliability levels, and State Road used 80% to 85% of reliability levels. Since a road with heavier traffic typically requires a higher level of reliability, the highest reliability levels were selected for the Interstate Highway and the lowest for State Road. The upper limit of reliability was used to determine the threshold between poor and fair conditions, while the lower limit reliability level was used to calculate the threshold between fair and good conditions.

The FWD data collected from the INDOT pavement management system (PMS) database was used for reliability analysis. The total number of FWD data points were 1579, 605, and 717, for State Road, US-Highway, and Interstate Highway, respectively. Empirical cumulative distribution function (ECDF) was used to conduct a reliability analysis for determining preliminary threshold values of FWD parameters. The ECDF is an estimate of the cumulative distribution of each data point in the measured samples, that allows to identify a value of FWD parameter, corresponding to the target percentage or reliability level. As an example, **Figure 2** shows the Interstate Highway ECDF plot for all FWD deflection parameters. For example, the D0 values corresponding to the upper and lower reliability levels were 214.9 and 149.1 microns, respectively. This means that the Interstate Highway is in poor condition, when the D0 is greater than 214.9 microns. The same approach was applied to other FWD parameters as shown in **Figure 2**, and the reliability results of FWD deflection parameters for each road classification are summarized in **Table 1**. The values of FWD parameter, corresponding to four reliability levels (80%, 85%, 90%, and 95%) are presented in **Table 1** to confirm that the FWD deflection parameter increases as the reliability level increases. Furthermore, it was found that the State Road exhibited the greatest values of FWD deflection parameters, while the Interstate Highway showed the lowest values at the same reliability level. This is a reasonable trend, because a conservative threshold value should be applied to the road with heavier traffic (i.e., Interstate Highway). Consequently, the preliminary threshold values for all FWD deflection parameters are highlighted in **Table 1**.

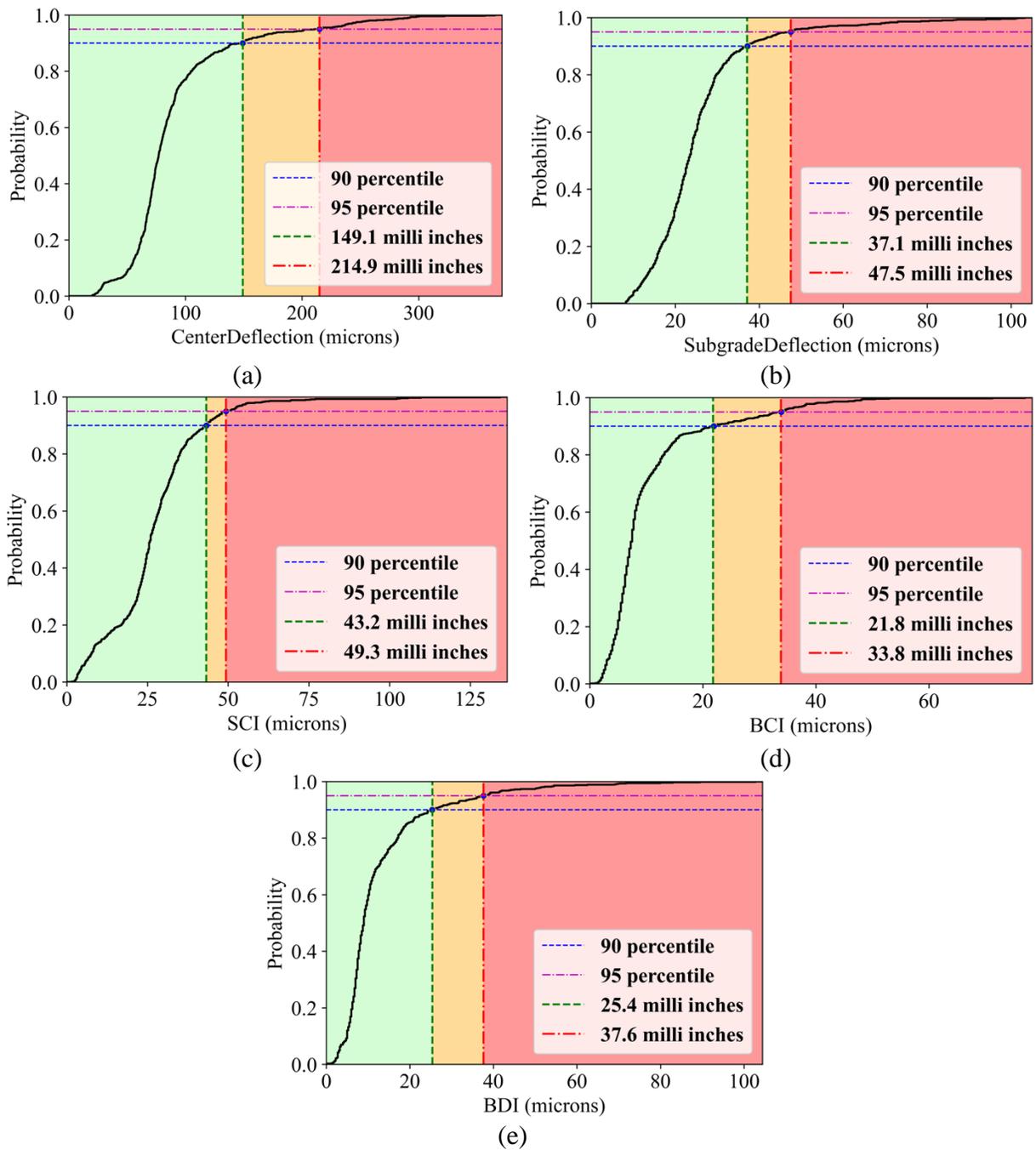

**Figure 2 ECDF plot for FWD deflection parameters on Interstate full depth asphalt pavement: (a) center deflection (D0), (b) subgrade deflection (D60), (c) SCI, (d) BCI, and (e) BDI**

**TABLE 1 Reliability Results of FWD Deflection Parameters**

| Road Type | Percentage (%) | D0 (microns) | D60 (microns) | SCI (microns) | BCI (microns) | BDI (microns) |
|---|---|---|---|---|---|---|
| *SI Unit* | | | | | | |
| State Road | 80 | 359.9 | 59.4 | 111.0 | 57.2 | 81.5 |
| | 85 | 388.6 | 62.2 | 123.2 | 62.2 | 89.2 |
| | 90 | 423.7 | 67.3 | 138.7 | 68.1 | 102.9 |
| | 95 | 494.3 | 77.5 | 171.2 | 82.6 | 125.0 |
| US-Highway | 80 | 212.9 | 50.5 | 58.9 | 30.5 | 45.5 |
| | 85 | 227.6 | 53.1 | 66.0 | 34.3 | 50.0 |
| | 90 | 259.8 | 56.6 | 76.7 | 39.1 | 55.9 |
| | 95 | 295.4 | 63.0 | 94.0 | 46.2 | 68.6 |
| Interstate Highway | 80 | 106.4 | 30.2 | 35.6 | 13.2 | 17.3 |
| | 85 | 121.4 | 33.0 | 38.1 | 15.5 | 19.6 |
| | 90 | 149.1 | 37.1 | 43.2 | 21.8 | 25.4 |
| | 95 | 214.9 | 47.5 | 49.3 | 33.8 | 37.6 |

| Road Type | Percentage (%) | D0 (mils) | D60 (mils) | SCI (mils) | BCI (mils) | BDI (mils) |
|---|---|---|---|---|---|---|
| *US Unit* | | | | | | |
| State Road | 80 | 14.2 | 2.3 | 4.4 | 2.3 | 3.2 |
| | 85 | 15.3 | 2.5 | 4.9 | 2.5 | 3.5 |
| | 90 | 16.7 | 2.7 | 5.5 | 2.7 | 4.1 |
| | 95 | 19.5 | 3.1 | 6.7 | 3.3 | 4.9 |
| US-Highway | 80 | 8.4 | 2.0 | 2.3 | 1.2 | 1.8 |
| | 85 | 9.0 | 2.1 | 2.6 | 1.4 | 2.0 |
| | 90 | 10.2 | 2.2 | 3.0 | 1.5 | 2.2 |
| | 95 | 11.6 | 2.5 | 3.7 | 1.8 | 2.7 |
| Interstate Highway | 80 | 4.2 | 1.2 | 1.4 | 0.5 | 0.7 |
| | 85 | 4.8 | 1.3 | 1.5 | 0.6 | 0.8 |
| | 90 | 5.9 | 1.5 | 1.7 | 0.9 | 1.0 |
| | 95 | 8.5 | 1.9 | 1.9 | 1.3 | 1.5 |

*Verification of FWD Thresholds Using the Structural Number Ratio (SNR)*

SNR is derived from the concept of a pavement structural number developed by the American Association of State Highway Officials (AASHO) and it has been widely used in PMS to evaluate the structural capacity of in-service asphalt pavements *(27, 28)*. As expressed in **Equation 1**, the SNR is the ratio of the effective structural number ($SN_{eff}$) and the required structural number ($SN_{req}$) *(17, 29)*. The $SN_{eff}$ indicates the structural number of existing asphalt pavements, and the $SN_{req}$ is the minimum structural number to ensure a desirable structural performance over the design period. Theoretically, the $SN_{eff}$ should be greater than the $SN_{req}$ to achieve an adequate structural performance, and the minimum SNR requirement is unity.

$$SNR = \frac{SN_{eff}}{SN_{req}} \tag{1}$$

The American Association of State Highway and Transportation Officials (AASHTO) 1993 pavement design guide dictates the methods to determine the $SN_{eff}$ and $SN_{req}$ using the FWD deflection data *(26)*. The $SN_{req}$ can be determined by solving the **Equation 2**, and input parameters are a subgrade resilient modulus ($M_R$), equivalent single axle load (ESAL), reduction in serviceability ($\Delta PSI$), standard deviation ($S_0$), and standard normal deviate ($Z_R$). In this study, the FWD deflection measured at 1524 mm (60 in.) away from the loading center was used to calculate the $M_R$. The same ESAL was assumed for each road classification, based on the ESAL category dictated in the INDOT specification. The State Road used one million ESALs, while US-Highway and Interstate Highway used four million and 10 million ESALs, respectively. In addition, 1.701 of $\Delta PSI$ and 0.35 of $S_0$, which are recommended in the INDOT specification, were used. The $Z_R$ was determined based on the reliability levels following the AASHTO 1993 design guide: -1.037 for 85% reliability level, -1.282 for 90% reliability level, -1.645 for 95% reliability level. It should be noted that the same reliability levels used for the reliability analysis were applied to the $SN_{req}$ calculation (i.e., State Road: 85%, US-Highway: 90%, and Interstate Highway: 95%).

$$\log ESAL = Z_R \times S_0 + 9.36 \times \log(SN_{req} + 1) - 0.2 + \frac{\log\left(\frac{\Delta PSI}{4.2 - 1.5}\right)}{0.4 + \frac{1094}{(SN_{req} + 1)^{5.19}}} + 2.32 \times \log M_R$$
$$- 8.07 \tag{2}$$

$$M_R = \frac{0.24 \times P}{d_r \times r}$$

where, ESAL is the equivalent single axle load, $\Delta PSI$ is a reduction in serviceability, $S_0$ is a standard deviation, $Z_R$ is a standard normal deviate, $M_R$ is a subgrade resilient modulus (ksi), P is a FWD load (lbs.), r is a distance from the center of load (in.), and $d_r$ is a deflection at a distance r from the center of the load (in.).

According to the AASHTO 1993 design guide, the $SN_{eff}$ can be determined using a total pavement thickness and the effective modulus of existing pavements obtained from the FWD data. However, the AASHTO 1993 method requires a trial-and-error procedure to obtain the pavement effective modulus, which is a key parameter for $SN_{eff}$ calculation, and is impractical to be incorporated into the network level PMS. Furthermore, the AASHTO 1993 method is significantly dependent on the pavement thickness, resulting in overestimated $SN_{eff}$. Thus, previous researchers have developed alternative models to estimate the $SN_{eff}$ using the FWD deflection data, without a trial-and-error procedure *(9, 17, 30)*. Recently, a new $SN_{eff}$ prediction model was developed and verified with field data in a concurrent study conducted by INDOT. As expressed in **Equation 3**, the new model only requires a total pavement thickness and the FWD deflection basin parameter, the area under pavement profile (AUPP), to make accurate predictions of $SN_{eff}$. Therefore, this study adopts this new model to calculate the $SN_{eff}$.

$$SN_{eff} = 2.272 \times H_p^{0.4217} \times AUPP^{-0.4678} \tag{3}$$

where, $H_p$ is the pavement thickness above subgrade (in.), and AUPP is the area under pavement profile (mils).

The maximum FWD deflection (D0) was selected as a representative FWD parameter to evaluate the reliability-based approach and to verify the preliminary threshold values. Since both D0 and SNR represent the overall structural conditions of existing pavements, the D0 was the most appropriate FWD parameter to compare with the SNR. Other FWD parameters are typically used to estimate the structural conditions of specific layers (i.e., surface layer, base layer, and subgrade), indicating that a more specific structural index may be needed to further verify the other FWD parameters, other than the SNR. However, there are currently no available structural indices for the specific layers. Therefore, it was assumed that the reliability-based approach verified with D0 can be extended to the other FWD parameters to determine the final threshold values.

**Figure 3** shows that the SNR decreased as D0 increased with an exponential relationship, for all road classifications, including State Road, US-Highway, and Interstate Highway. Based on the relationship, the D0 values, corresponding to the SNR limit were identified to establish the D0 threshold range for each road classification. As shown in **Figure 3**, the D0 threshold of State Road ranged from 125 to 400 microns, and the D0 threshold range was identified as 120 to 254 microns for US-Highway and 100 to 203 microns for Interstate Highway, respectively. Overall, the threshold of State Road was greater than the threshold of Interstate Highway. This trend is consistent with that of reliability-based threshold values: a road with heavier traffic requires a lower level of threshold to be conservative. Interestingly, for all road classifications, the upper limit of threshold range was almost identical to the preliminary upper limit D0 threshold value from the reliability analysis (in **Table 1**) to distinguish between poor and fair conditions. For example, the upper limit of reliability-based threshold D0 for State Road (388.6 microns) was slightly smaller than the upper limit of D0 threshold identified based on the SNR analysis (400 microns). Therefore, the relationship between SNR and D0 verified that the reliability-based D0 threshold values are acceptable to distinguish good, fair, and poor structural conditions of full-depth asphalt pavements, and the final threshold values of FWD deflection parameters are presented in **Table 2**.

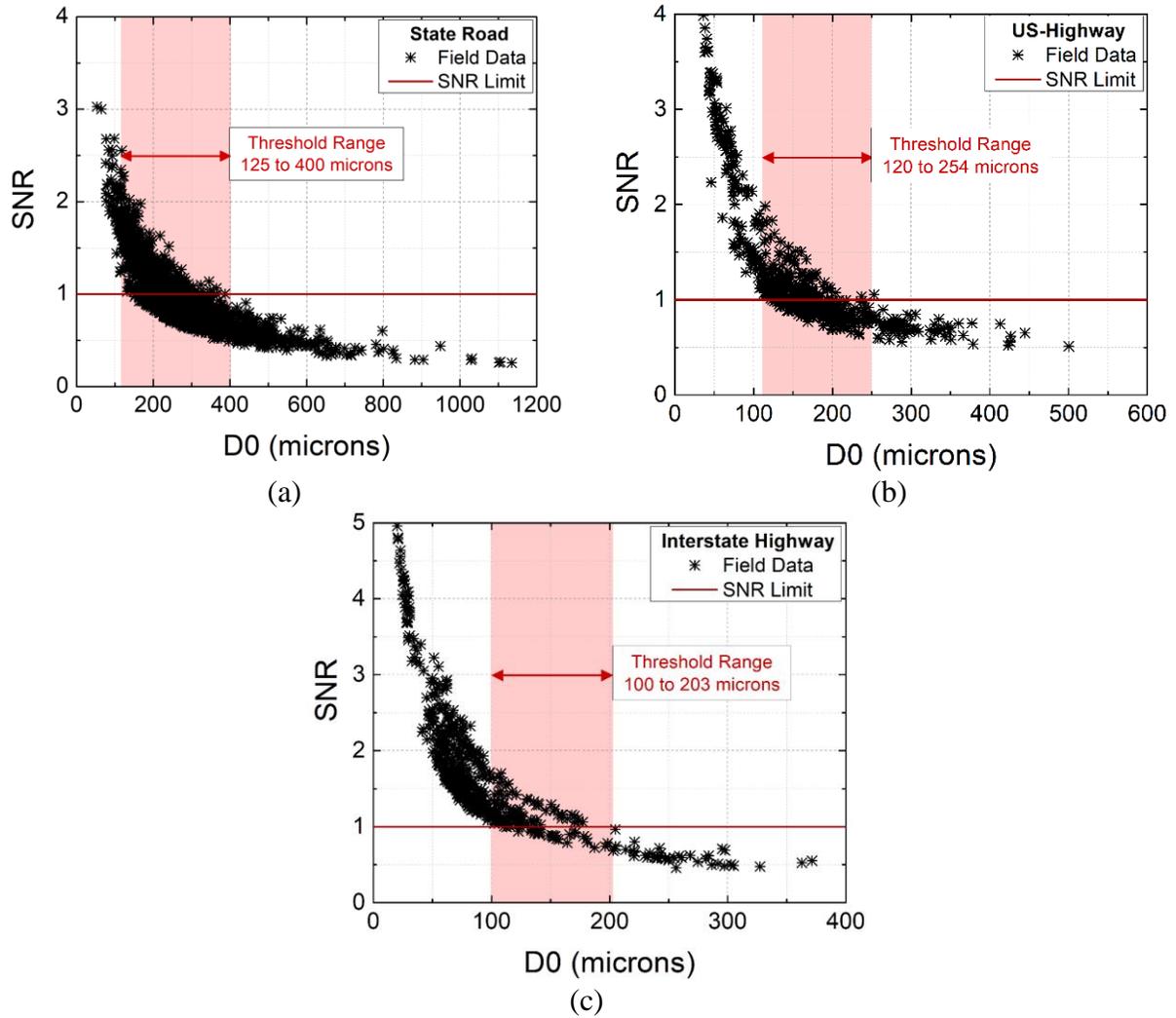

(a)

(b)

(c)

**Figure 3 Comparison between SNR and D0: (a) State Road, (b) US-Highway, and (c) Interstate Highway**

**TABLE 2 Final Threshold Values of FWD Deflection Parameters**

| Road Type | FWD Parameter | Good | Fair | Poor |
|---|---|---|---|---|
| | | *SI Unit* | | |
| | D0 (microns) | D0 < 359.9 | 359.9 < D0 < 388.6 | 388.6 < D0 |
| | D60 (microns) | D60 < 59.4 | 59.4 < D60 < 62.2 | 62.2 < D60 |
| State Road | SCI (microns) | SCI < 111.0 | 111.0 < SCI < 123.2 | 123.2 < SCI |
| | BCI (microns) | BCI < 57.2 | 57.2 < BCI < 62.2 | 62.2 < BCI |
| | BDI (microns) | BDI < 81.5 | 81.5 < BDI < 89.2 | 89.2 < BDI |
| | D0 (microns) | D0 < 227.6 | 227.6 < D0 < 259.8 | 259.8 < D0 |
| | D60 (microns) | D60 < 53.1 | 53.1 < D60 < 56.6 | 56.6 < D60 |
| US-Highway | SCI (microns) | SCI < 66.0 | 66.0 < SCI < 76.7 | 76.7 < SCI |
| | BCI (microns) | BCI < 34.3 | 34.3 < BCI < 39.1 | 39.1 < BCI |
| | BDI (microns) | BDI < 50.0 | 50.0 < BDI < 55.9 | 55.9 < BDI |
| | D0 (microns) | D0 < 149.1 | 149.1 < D0 < 214.9 | 214.9 < D0 |
| | D60 (microns) | D60 < 37.1 | 37.1 < D60 < 47.5 | 47.5 < D60 |
| Interstate Highway | SCI (microns) | SCI < 43.2 | 43.2 < SCI < 49.3 | 49.3 < SCI |
| | BCI (microns) | BCI < 21.8 | 21.8 < BCI < 33.8 | 33.8 < BCI |
| | BDI (microns) | BDI < 25.4 | 25.4 < BDI < 37.6 | 37.6 < BDI |
| | | *US Unit* | | |
| | D0 (mils) | D0 < 14.2 | 14.2 < D0 < 15.3 | 15.3 < D0 |
| | D60 (mils) | D60 < 2.3 | 2.3 < D60 < 2.5 | 2.5 < D60 |
| State Road | SCI (mils) | SCI < 4.4 | 4.4 < SCI < 4.9 | 4.9 < SCI |
| | BCI (mils) | BCI < 2.3 | 2.3 < BCI < 2.5 | 2.5 < BCI |
| | BDI (mils) | BDI < 3.2 | 3.2 < BDI < 3.5 | 3.5 < BDI |
| | D0 (mils) | D0 < 9.0 | 9.0 < D0 < 10.2 | 10.2 < D0 |
| | D60 (mils) | D60 < 2.1 | 2.1 < D60 < 2.2 | 2.2 < D60 |
| US-Highway | SCI (mils) | SCI < 2.6 | 2.6 < SCI < 3.0 | 3.0 < SCI |
| | BCI (mils) | BCI < 1.4 | 1.4 < BCI < 1.5 | 1.5 < BCI |
| | BDI (mils) | BDI < 2.0 | 2.0 < BDI < 2.2 | 2.2 < BDI |
| | D0 (mils) | D0 < 5.9 | 5.9 < D0 < 8.5 | 8.5 < D0 |
| | D60 (mils) | D60 < 1.5 | 1.5 < D60 < 1.9 | 1.9 < D60 |
| Interstate Highway | SCI (mils) | SCI < 1.7 | 1.7 < SCI < 1.9 | 1.9 < SCI |
| | BCI (mils) | BCI < 0.9 | 0.9 < BCI < 1.3 | 1.3 < BCI |
| | BDI (mils) | BDI < 1.0 | 1.0 < BDI < 1.5 | 1.5 < BDI |

**Determination of Functional Parameters Thresholds**

Thresholds of IRI and CD were determined using the reliability percentile values verified by the threshold comparison results of FWD deflection parameters. **Figure 4** shows ECDF plots of IRI and CD for Interstate Highway. The 95% and 90% reliability levels were used for rating the pavements in good, fair, and poor condition. Accordingly, the higher IRI threshold value was 2.07 m/km, and the lower IRI threshold value was 1.73 m/km. Similarly, the higher threshold of CD was 13.2%, while the lower threshold value for CD calculation was 12.5%. It is important to note that thresholds of IRI and CD were determined only for full depth asphalt pavement,

Interstate Highway, and IRI and CD data collection is ongoing for State Road and US-Highways. Thus, in this study, Interstate Highway was used to verify the concept and developed algorithm of patching suggestions.

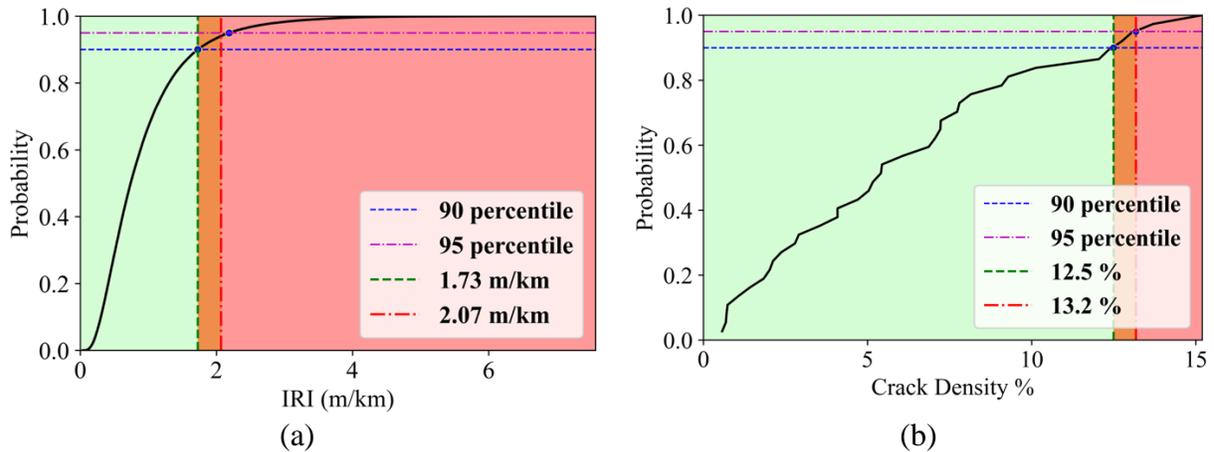

(a)                                                     (b)

**Figure 4 ECDF plot for functional parameters of Interstate Highway: (a) IRI and (b) CD**

## EVALUATION OF PAVEMENT CONDITION RATING PARAMETERS
### Comparison of Overall Road Conditions
Interstate Highway I-70 was used as a representative full-depth asphalt pavement to determine whether the PCR parameters can be used to assess overall road conditions. The overall conditions of driving lane (DL) and passing lane (PL) for both east-bound (EB) and west-bound (WB) roads were compared using box plots. **Figure 5** shows a comparison of IRI values on driving and passing lanes, based on the box plot parameters. Here, the bottom edge of the rectangle denotes the first quantile (25%) mark, and the upper edge of the rectangle denotes the third quantile (75%) mark in the PCR parameter measurement. The meaning of this data is represented by the green triangle and median is represented by the green horizontal line. The length of the vertical line denotes the range of the data while the minimum and maximum values are shown at the edges of this line. The point markers represent the PCR parameter measurements above the threshold of each road section. Overall, this box plot analysis result shows that both right wheel IRI and left wheel IRI values of driving lane are greater than those in the passing lane for both EB and WB. Usually, a driving lane has more wear than a passing lane, due to higher traffic and lower vehicle speed on the driving lane. This indicates that the functional parameters can correctly compare the overall functional condition of lanes and detect roads which require attention.

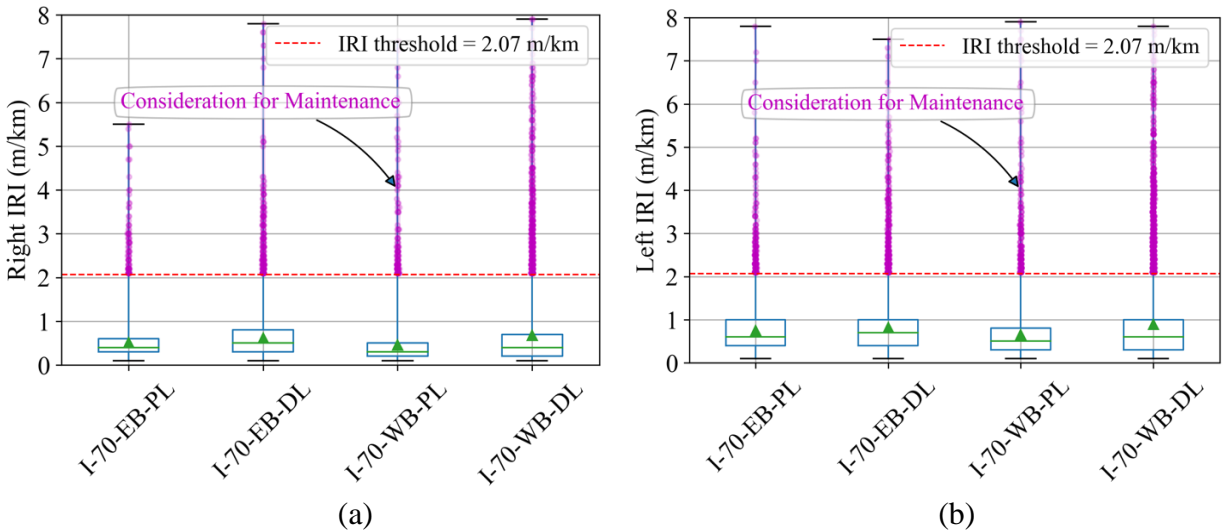

(a)                                                                              (b)

**Figure 5 Comparison of IRI on I-70: (a) Right IRI and (b) Left IRI**

As shown in **Figures 6a**, and **6b**, a similar trend was observed from D0 and D60 on the same road section on I-70. Both D0 and D60 values of driving lane were clearly greater than the passing lane based on the box plot parameters. As expected, driving lane results exhibited (i) higher variation in both D0 and D60 represented by a bigger box size and (ii) significantly higher deflection values, compared to passing lane results. Higher density of traffic on driving lanes exacerbates the pavement stress causing greater deflection values and consequently increased variation in distribution of D0 and D60. In addition to the overall road/lane comparison, **Figure 6c** draws the comparison of FWD deflection parameters across the road classifications. Since Interstate Highway is typically designed with stronger pavement structures to resist heavier traffic, overall FWD deflection parameters are lower than on US-Highway and State Road. This means that the FWD deflection parameters can be used as structural indicators to represent overall road conditions.

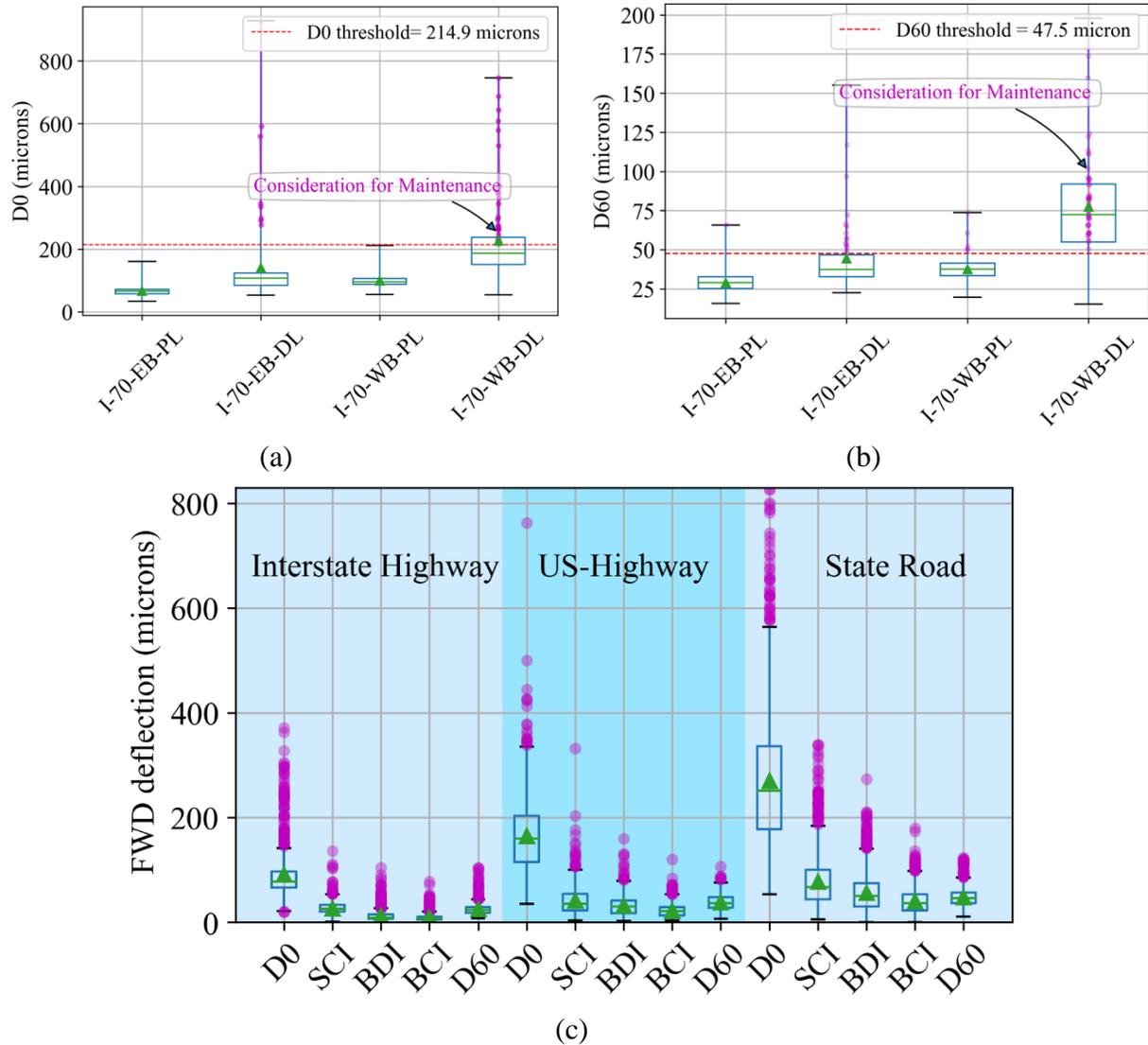

(a)                                                          (b)

(c)

**Figure 6 Comparison of FWD parameters: (a) D0 of I-70, (b) D60 of I-70, and (c) All FWD deflection parameters from all road classifications**

**Identifying Localized Distresses Using Pavement Condition Rating Parameters**
Identifying localized distresses is the major expectation from the PCR parameters in providing accurate geo-location of road sections with maintenance requirements. Spatially integrated IRI and FWD deflection parameters were compared with the crack density, road surface and ROW images to determine whether these parameters can successfully capture the localized distresses of in-service full-depth asphalt pavements. A stem plot comparison of the three pavement condition indices at each DMI number for the same section of I-70 as in previous section is shown in **Figure 7**. The greatest value of L-IRI 8.1 m/km was observed at the left wheel path, DMI 2924, on I-70-WB-PL as shown in the plot in **Figure 7a**. This value is also greater than the interstate IRI threshold value. The road surface and ROW image of the same DMI number 2924 captured from the WayLink software shows a localized patching for pothole on the left edge of the road, marked with a red arrow in **Figure 7a**.

Another example shows that strip patching was captured at DMI number 330 on the same section of I-70 in **Figure 7b**. Both left and right IRI values; R-IRI (7.7 m/km) and L-IRI (5.4 m/km) were also greater than the IRI threshold value of 2.07 m/km as shown **Figure 7b**. The authors extensively checked the ability of the 1.8-m resolution data to detect localized distresses by selecting assessment index values exceeding their upper threshold values from scatter plots and comparing them to their corresponding ROW image and road surface image. These findings confirm that the selected PCR parameters accurately detected localized distresses and other features on roads such as crack sealing, small patching, uphill, downhill, turns, railroad crossings etc. from road condition parameter data alone.

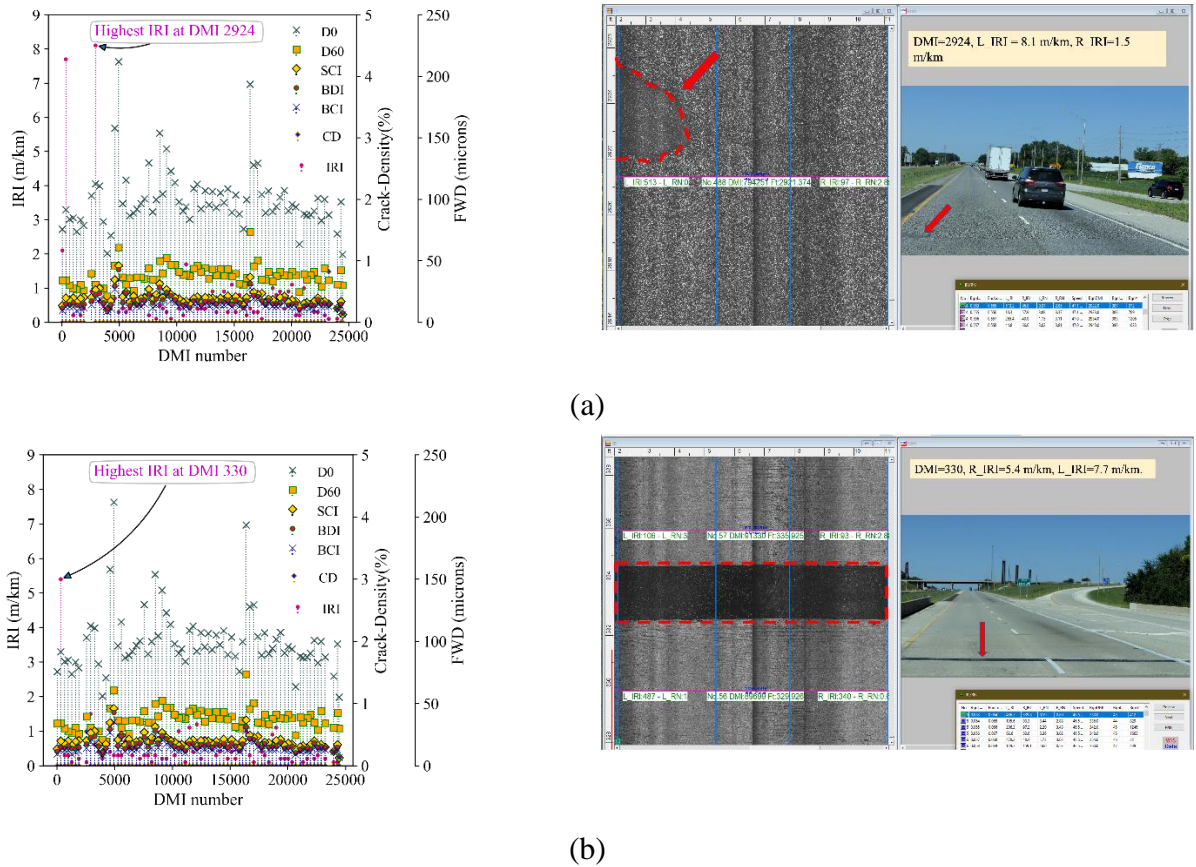

(a)

(b)

**Figure 7 Stem plot for IRI, FWD deflections, and CD, along with the ROW image and road surface image of DMI: (a) at DMI 2924, and (b) at DMI 330 on I-70.**

## DEVELOPMENT OF PATCHING MANAGER TOOL (PMT)

### Patching Suggestion Algorithm

Pavement surface patching and full depth patching were used as the patching depth suggestions in the surface condition evaluation. Surface patching and full depth patching are defined according to the AASHTO pavement maintenance protocol. Patching required and patching warning were similarly used as patching priority suggestion. Patching required has higher priority in maintenance than patching warning. **Figure 8** illustrates the classification of the state of the three pavement layers into five conditions for patching. The underlying rectangular shapes represent the surface, base/intermediate and sub- base/subgrade layers of a full-depth asphalt pavement. The red, orange, and green colored layers represent the current state of the layers.

Green represents the good state of layers i.e., values lower than the lower thresholds of PCRs, orange represents the fair state, and red represents the poor state of pavement layers i.e., PCR values greater than the upper thresholds calculated earlier.

*Condition 1* shows the poor condition of the subbase/subgrade layer with a combination of poor state of the intermediate/base layer. Pavement in this state requires a full-depth patch irrespective of the condition of the surface layer. Similarly, when the intermediate/base layer is in a poor state and subbase/subgrade layer is in fair state, *Condition 2* would be applicable and a warning for full-depth patching would be suggested by the algorithm. Furthermore, the patching suggestion requirement and warning for surface patching are based on two combinations of pavement layer states as shown in *Conditions 3* and *4*. *Conditions 3* requires surface layer patching determined by poor condition of surface layer with fair state of base/intermediate layer and good condition of subbase/subgrade layer. *Condition 4* indicates warning surface layer patching determined by fair state of surface layer and good conditions of base/intermediate and subbase/subgrade layers. Lastly, *Condition 5* indicates good road condition where no further action is needed. This conditional patching suggestion algorithm is reinforced by adding pavement type/road classification threshold values. The threshold-based patching suggestion for Interstate Highway full-depth asphalt pavement is aggregated in **Table 3**.

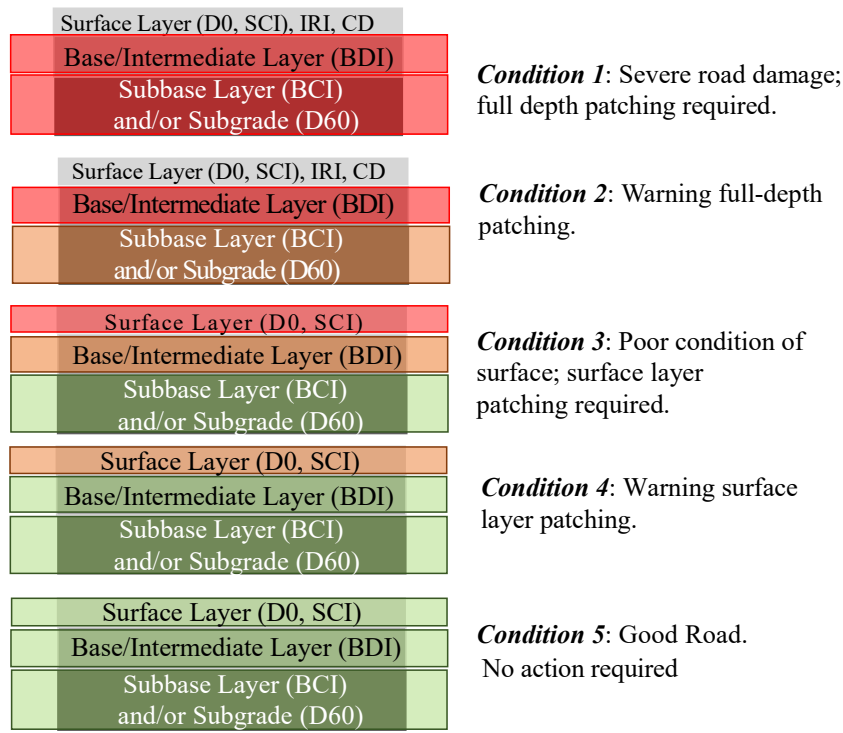

**Figure 8  Patching conditions labeled according to the measurement range of FWD deflections**

**Table 3 Threshold-Based Patching Suggestion Algorithm for Interstate Full-Depth Asphalt Pavement**

| IRI (m/km) | CD (%) | D0 (µm) | SCI (µm) | BDI (µm) | D60 (µm) | BCI (µm) | Patching Suggestions |
|---|---|---|---|---|---|---|---|
| - | - | - | - | - | > 47.5 | >33.8 | Severe road damage; full-depth patching requirement |
| | | | | | > 37.1 & ≤ 47.5 | > 21.8 & ≤ 33.8 | Full-depth patching warning |
| > 2.07 | >13.2 | >214.9 | > 49.3 | > 37.6 | ≤ 37.1 | ≤ 21.8 | Poor condition of surface layer; surface patching required |
| > 1.73 & ≤2.07 | >12.5 & ≤13.2 | > 149.1 & ≤ 214.9 | >43.2 & ≤49.3 | > 25.4 & ≤ 37.6 | ≤ 37.1 | ≤ 21.8 | Surface patching warning |
| ≤ 1.73 | ≤12.5 | ≤149.1 | ≤ 43.2 | ≤ 25.4 | ≤ 37.1 | ≤ 21.8 | No action required |

**Web-Based Application: PMT**

A web-based application can be hosted on an internet browser and hence is easily available to view with credentials. The open-source structure of the application also makes it free to use for individual analysts/engineers. In the PMT application, the threshold-based patching algorithm is currently implemented using the spatially indexed and integrated PCR parameters to create patching suggestions for roads of interest. The patching suggestions can be downloaded into patching tables with location co-ordinates, PCR parameter measurement values, suggested parameter severity ratings, patching information (i.e., area, depth, quantity and priority suggestion), and pavement information (i.e., road name, pavement type, reference point (RP) and lane). **Figure 9** illustrates a block diagram and thresholds-based patching suggestion table for PMT.

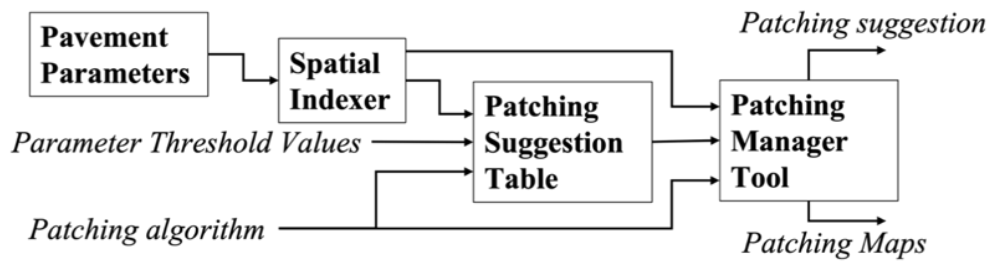

**Figure 9 Block diagram for the patching manager tool.**

The developed web- application has three major functionalities. Once a road of interest is selected, it creates suggestions for patching location, quantity, and a color-coded visualization of the patching suggestions. The visualization shows markers which represent a pavement section of 1.8 m length and 3.6 m width, as shown in **Figure 10**. These markers are identified by the unique DMI number and geographical coordinates. The marker color represents patching depth and priority. The patching suggestion feature also includes histogram analysis and scatter plot analysis of pavement rating parameters. These analysis tools are necessary to analyze the values at the location nearest to a distress to understand the probable causes of distress. The cause of distress is an important factor in determining the maintenance suggestion. The potential of the web application is placed on its data-driven ability to allow users to view road conditions at distress locations in a three-dimensional space and can compare them in real time with other completely different locations hundreds of miles apart.

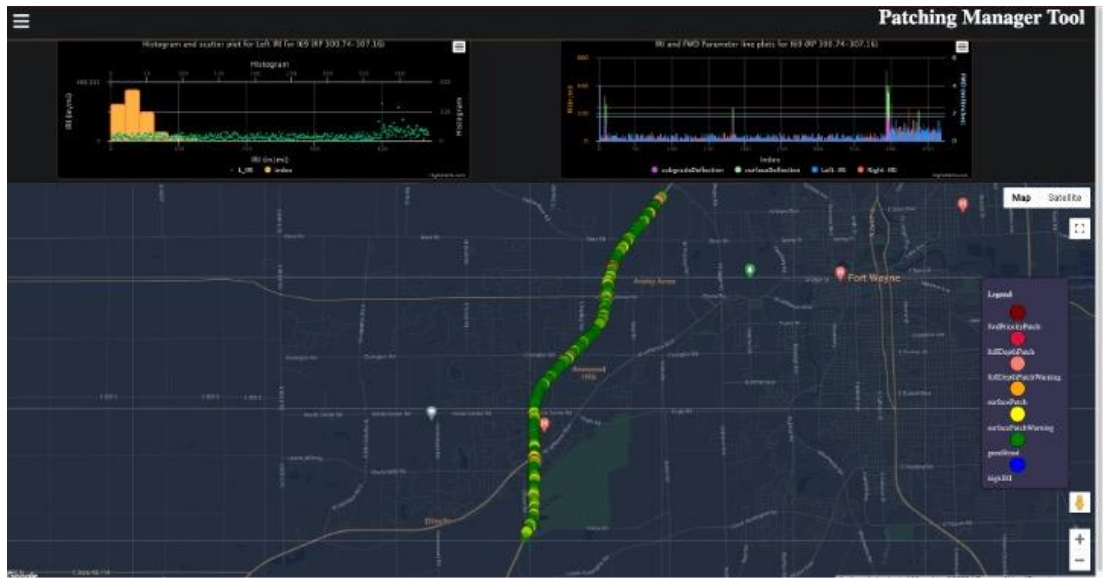

(a)

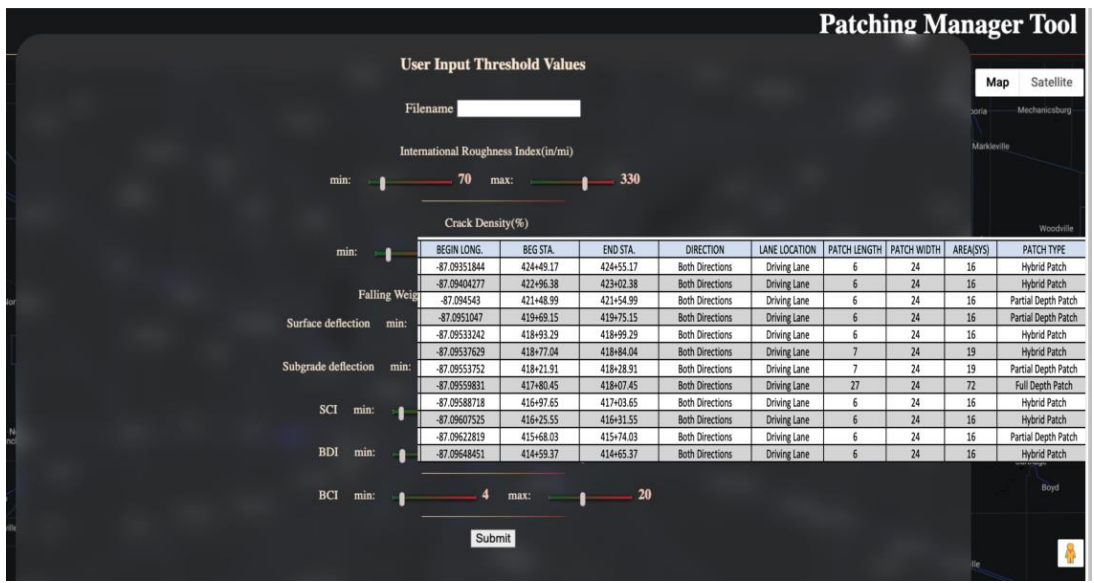

(b)

**Figure 10 Web-based patching manager tool: (a) Patching suggestion for the example road (I-69) using color coded markers and graphs for analysis of parameters, and (b) User input threshold selector for patching suggestion calculation.**

To verify the developed web based PMT application, the patching suggestions provided from PMT were compared with the Google Street View feature in PMT. An example of road (I-64) was used for this verification. A particular DMI 24204 on I-64 EB DL is shown by a yellow dot in **Figure 11**, which was selected from the patching suggestion visualization, and street view at this DMI was explored. PMT suggests surface patching warning for the DMI with a yellow dot, because the IRI at this location is greater than its lower-level threshold value. Therefore, it was confirmed that PMT accurately suggests patching suggestion at the distressed location captured in the Google Street View map.

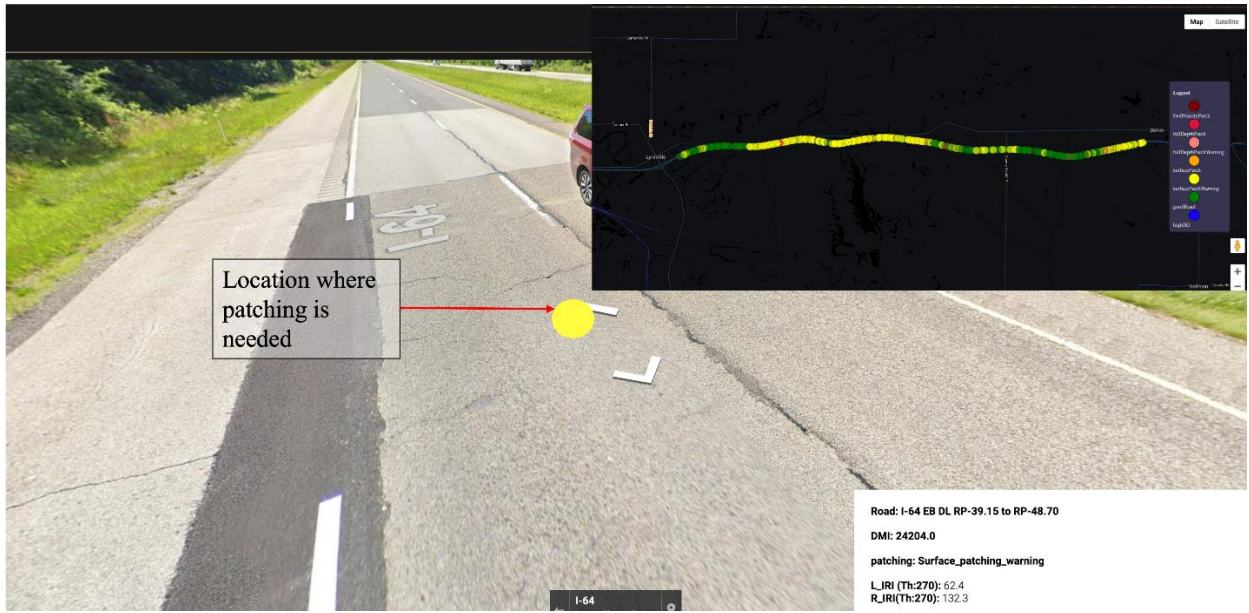

**Figure 11  Example of final output of patching suggestion in PMT**

**CONCLUSION**
This study proposes a reliability-SNR-based threshold calculation method for integrated PCR parameters. The data fusion approach integrating structural and functional assessment parameters with the matched geographical location describes the overall pavement assessment with better accuracy. It considers roughness, cracking and FWD deflections at the same location, allowing the automation of improved patching suggestions. The patching suggestion algorithm would incorporate the overall pavement condition information and could create five pavement condition ratings based on the distress level in three major layers of pavement structure. The algorithm was tested using calculated PCR parameter threshold values for Interstate full-depth asphalt pavement segments. Comparative analysis of PCR parameters was able to detect localized distresses. It could reduce the bulk human resource and time required in manual routine survey. The developed threshold-based patching suggestion algorithm was reinforced with a web-based application: PMT to improve current practice of PMS. The key features of PMT web-based application are summarized below:

- The application plots comprehensive patching suggestions with color-coded markers for a selected road segment on a map.

- PMT integrates two-dimensional road surface and ROW images to aid visual field assessments.
- PMT allows users to take advantage of Google Street View at the patching locations for 3D virtual spatial analysis.
- Histogram and scatter plots are included to analyze distribution or pattern in PCR parameters along/among the selected roads.
- PMT supports manual adjustment of threshold values for each PCR parameter in suggestion algorithm. This feature creates flexibility for users to customize the patching tables, which are essential for analyzing and updating maintenance standards.

The PMT web-based application can provide the high-resolution comprehensive patching suggestion with key features summarized above, and it was successfully verified with ROW images and Google Street View map. Furthermore, this study introduced the approach to determine appropriate threshold values based on local road conditions, and this approach can be used for roads in other states or countries. Consequently, the PMT web-based application, along with the proposed approach for thresholds determination can be applied to any region with different road conditions.

As a next step, ground penetrating radar (GPR) and traffic speed deflectometer (TSD) will be included as additional PCR parameters to improve the accuracy and increase the resolution of patching depth suggestions. In addition, the database structure in PMT web-based application will be scaled up to display and analyze larger datasets like a statewide road network on map simultaneously

## ACKNOWLEDGMENTS

This work was supported in part by the Joint Transportation Research Program administered by the Indiana Department of Transportation and Purdue University. This paper's contents reflect the views of the authors, who are responsible for the facts and accuracy of the data presented herein and do not necessarily reflect the official views or policies of the sponsoring organizations. These contents do not constitute a standard, specification, or regulation. We would also like to thank Professor Kelvin Wang and Dr Guangwei Yang at Oklahoma State University for assisting with WayLink systems corporation 3D imaging software.

## AUTHOR CONTRIBUTION STATEMENT

The authors confirm contribution to the paper as follows: Research idea conception development and design: Seonghwan Cho, James V. Krogmeier, and John E. Haddock; data processing, data integration and reliability data analysis: Sneha Jha and Yaguang Zhang, data collection: Bongsuk Park; analysis and results interpretation: Seonghwan Cho, James V. Krogmeier, John E. Haddock, Sneha Jha, Yaguang Zhang and Bongsuk Park; literature review: Sneha Jha and Tandra Bagchi; draft manuscript preparation and reviews: All authors. All authors reviewed the results and approved the final version of the manuscript.